\documentclass[twocolumn,showpacs,preprintnumbers]{revtex4}

\usepackage{graphicx}
\usepackage{psfig}
\usepackage{dcolumn}
\usepackage{bm}
\def \frac#1#2{ { #1 \over #2} }

\begin{document}


\title{Quantum calculation of vortices in  the inner crust of neutron stars.}

\author{P. Avogadro$^{a,b}$, F. Barranco$^{c}$, R.A. Broglia$^{a,b,d}$, 
E. Vigezzi$^{b}$\\
$^a$ Dipartimento di Fisica, Universit\`a degli Studi di Milano,
via Celoria 16, 20133 Milano, Italy.\\
$^b$ INFN, Sezione di Milano, via Celoria 16, 20133 Milano, Italy.\\
$^c$ Departamento de Fisica Aplicada III, Escuela Superior de Ingenieros, Camino 
de los Descubrimientos s/n,
41092 Sevilla, Spain.\\
$^d$ The Niels Bohr Institute, University of Copenhagen, Blegdamsvej 17, 2100 
Copenhagen \O, Denmark.}

\date{\today}



\begin{abstract}
We study, within a
quantum mechanical framework based on self-consistent mean field theory, the
interaction between a vortex  and a nucleus immersed in a sea of free neutrons,
a scenario representative of the inner crust of neutron stars. 
Quantal finite size effects  force the vortex core outside the nucleus,
influencing vortex pinning in an important way.

\end{abstract}

\maketitle
Neutron stars (pulsars)
usually rotate 
with such a precision that they are known as the best timekeepers in the
universe. 
But every so often their rotation rate increases \cite{Pines}. 
Anderson and Itoh  proposed 
that these glitches can be viewed 
as "vorticity jumps", equivalent to "flux jumps" in a superconducting magnet
\cite{Anderson}.

The density of the neutron star increases going from the surface to the interior, 
and  when it becomes larger than about $n=$ 5 $\times 10^{-4} $ fm $^{-3}$, 
it  becomes energetically favourable for
some of the neutrons and all of the protons to lump together in extremely 
neutron rich Sn-like nuclei (that is nuclei containing 40-50 protons and about
one hundred neutrons), surrounded by a sea of free neutrons. 
These nuclei form a pure electrostatic "Wigner"
lattice. As one goes deeper into the
inner crust, the 
lattice step decreases, 
going from about 90 fm at $n = 5 \times 10^{-4} $ fm$^{-3}$, to about 30 fm 
at  $n = 7 \times 10^{-2} $ fm $^{-3}$. 
There is strong evidence which testifies to the fact that  in this density range, the neutrons are superfluid
(cf. e.g.[3,4]).
It has been argued that the angular velocity of the superfluid in the inner
crust
of a neutron star changes either  by vortex creep or by vorticity jumps, the latter causing the
glitches. Although this scenario  has been investigated in a number of 
publications \cite{Bulgaclist}, there remain  many open questions associated with the variety
of approximations used in these studies. In an effort to shed light
into these questions, we report here the first fully quantal, 
self-consistent calculation of a  vortex, taking into account the  inhomogeneous character 
typical of the inner crust of a neutron star.

This allows one to calculate the  pinning energy, 
namely the difference between  the energy cost to create a vortex far from 
the nucleus and on top of it. 
Our calculations are performed for  a single Wigner cell,
considering a single nucleus, 
in keeping with previous estimates \cite{epstein}, which showed that the 
distance between two neighbouring nuclei of the Coulomb 
lattice  is large compared with  typical vortex dimensions, 
possibly except for the deepest layers of the inner crust.

The calculations were performed solving the Hartree-Fock-Bogoliubov (HFB) 
equations (often called the De Gennes equations), within a 
cylindrical box of radius 30 fm and height 40 fm, imposing that the wavefunctions
vanish at the border of the cell.
We have assumed that the vortex-nucleus system is axially symmetric
( with the vortex directed along the $z-$axis), 
but we have constrained the proton density distribution 
(associated with deeply bound states)  to have  spherical symmetry.

The De Gennes equations have the form
\begin{equation} 
\left( \matrix 
{
K + V  - E_F    &  \Delta \cr
 \Delta&  -(K + V - E_F) \cr   } \right )
\left ( \matrix {
U_{\alpha} \cr  
V_{\alpha} \cr}
\right )  = E_{\alpha}
\left ( \matrix {
U_{\alpha} \cr  
V_{\alpha} \cr}
\right ),
\end{equation}
and have to be solved simultaneously with the number equation. 

The kinetic energy operator (including the effective mass
associated with the Skyrme interaction) is denoted by $K$, 
$V(\rho,z)$ being the self-consistent 
Hartree-Fock mean field and $\Delta$  the ($S=0$)
pairing field.
It displays the functional form
\begin{equation} 
\Delta(\rho,z,\phi) = - g[n(\rho,z)] \sum_{\alpha} U_{\alpha}V^*_{\alpha} 
= \Delta(\rho,z) e^{i \nu \phi},                 
\end{equation}
where $\rho$ is the
distance to this axis in the $x-y$ plane and $\phi$ is the azimuthal angle,
while  $g$ denotes the density-dependent strength of a (contact) pairing 
interaction.

Equations (1)-(2) allow for different solutions,
which can be labeled by  the vortex
number $\nu$ ($\nu$ = 0,1,2...). 
We have mostly  considered the $\nu=1$ case, 
in which each Cooper pair carries one unit of angular momentum, its
projection along the $z-$axis being +1. 
Note that for $\nu=0$ one recovers the Negele-Vautherin
results \cite{Negvau}. 

The SkII force \cite{Skyrmev}     
was used for determining the HF  field, while in the pairing sector 
we have adopted the density-dependent contact 
interaction introduced in ref. \cite{Garrido}, corresponding to
the values $g[n]=-481(1-(n/n_o)^{0.45})$ MeV fm$^3$, 
where $n_o=0.08$ fm$^{-3}$.
This interaction reproduces the values 
of the  pairing gap calculated with the Gogny force in uniform neutron matter.

The $\phi$ dependence of the quasiparticle amplitudes $U,V$ is
\begin{equation}
U_{\alpha}(\rho,z,\phi)=U_{\alpha}(\rho,z) e^{i l_{\alpha}  \phi};
V_{\alpha}(\rho,z,\phi)=V_{\alpha}(\rho,z) e^{i [l_{\alpha} - \nu] \phi}. 
\end{equation}
The  quasiparticle amplitudes are expanded on a basis of 
(free) single-particle wavefunctions inside the cylinder
\begin{equation}
U_{\alpha}(\rho,z)= \sum_{nm} U_{\alpha}^{n,m} 
\psi_{n,l_{\alpha}}(\rho) \chi_{m}(z),
\end{equation}
and similarly for $V_{\alpha}$ (replacing $\psi_{n,l_{\alpha}}$ with
$\psi_{n,l_{\alpha}- \nu}$).
The functions
$\chi_{m}(z)$ are (longitudinal) plane waves and 
$\psi_{n,l_{\alpha}}(\rho)$ (radial) Bessel functions, associated with a
cylinder with perfecly reflecting  walls,
$l_{\alpha}$ being the single-particle angular momentum along the cylinder axis,
chosen as quantization axis.
To limit computational complexity we have not included the 
spin-orbit interaction term in the calculation of the single-particle levels. 
Concerning the protons, we have solved the $\nu=0$ equations 
in a spherical box of radius $R_p= 15 $ fm, 
using the same SkII force
as for the neutrons. 
The effect of the neutrons on the (deeply bound) protons, has been 
included after a spherical
average of the various neutron distributions has been carried out.

In what follows we discuss the results associated with the value 
$E_F=5.8$ MeV of the Fermi energy.
The spatial dependence of the (neutron) density $n$ and of the pairing gap
$\Delta$ for a nucleus immersed in the neutron sea is shown in Fig. 1. 
The results for the density  
essentially coincide with the results of Negele and Vautherin for the
corresponding (spherically symmetric) Wigner cell.
The radius $R$ of the nucleus is about 7.5 fm, the associated
diffusivity  being about 0.9 fm.
Far from the nucleus the value of the pairing gap is about 
$\Delta_{unif} \approx $2.2 MeV,  equal to the
value obtained in the case of uniform neutron matter 
with the adopted pairing interaction, while at
the nuclear surface $\Delta \approx$ 1 MeV. 
The strong suppression of the pairing gap inside the nucleus  is
a consequence of the density dependence of the pairing interaction which
reflects 
the behaviour of the $^1S_0$ phase shift as a function of the relative kinetic
energy of the pair of interacting nucleons. 

Let us  now study the modifications induced on $n$ and $\Delta$ by the presence of a
vortex. We first discuss the case of free neutrons, a system which mimics,
exception made at the edges of the cylinder, uniform neutron matter.
In Fig. 3(a)    
we display the associated pairing gap for the case of a
$\nu=1$ vortex. It is seen that $\Delta$ vanishes along the $z$-axis.
For small values of $\rho, \rho < 3-4$ fm, the gap increases linearly
as a function of $\rho$. 
Defining the vortex core as the value of
$\rho$ for which $\Delta(\rho_{core}) = \Delta_{unif}/2$, one obtains 
$\rho_{core} \approx 2 $ fm, a value which is similar to
that of the correlation length 
$\xi \approx \hbar v_F /2\Delta (\approx 5$ fm).
For larger values of  $\rho$ the gap increases more slowly, 
gradually approaching the value 
$\Delta_{unif}$.

Concerning the density  distribution (see Fig. 2(a)), 
an axially symmetric depletion 
around the vortex axis is observed, with a radius (defined as the value of $n$
at half saturation density) of the order of 3 fm, which
is also of the order of the average distance between particles ($r_s \approx 
1.92/k_F \approx 1.02 /(E_F/20)^{1/2}$ fm $\approx$ 3.6 fm).
The above results are similar to those found in ref. \cite{Bulgac} within the
framework of nuclear energy density functional, taking into account the
different asymptotic values of the pairing gap (cf. also \cite{elgaroy}). 

We now turn our attention to the case of a $\nu=1$ vortex pinned on the nucleus. The
associated neutron density and pairing gap are shown in Fig. 2(b)
and 3(b). Comparing with the results displayed in Figs. 2(a) and  3(a) 
respectively, it is seen that 
both the density depletion around the vortex axis  and the vortex
core are strongly influenced by the presence of the nucleus. In fact, the vortex is seen to skate on the surface 
of the nucleus displaying a small penetration length. In other words, the vortex
is essentially expelled by the nucleus.
The pairing gap at the surface of the nucleus
is strongly suppressed,
both compared to the case of a isolated nucleus in the absence of the vortex
(cf. Fig. 1(b)) and compared to the case of a vortex in uniform (neutron) matter
(cf. Fig. 3(a)). 
In particular, at $\rho \approx 7$ fm (and $z=$0) $\Delta$ displays a value 
of about 0.5 MeV (Fig. 3(b)), 1 MeV (Fig. 1(b)) and of 2 MeV (Fig. 3(a)),
respectively.
Furthermore, the presence of the nucleus delays the rise of the
pairing gap to its asymptotic value by 5-7 fm
as compared to the uniform
case  (for example, in the $z=0$ plane,
$\Delta $ becomes equal to 1.75 MeV at $\rho\approx $ 11 fm, compared
to $\rho= 6 $ fm in the uniform case). 

There are two contrasting effects which influence the behaviour of a $\nu=1$ vortex
in the presence of a nucleus and lead to the results discussed above.
First,  the small (large) value of
the pairing gap inside (outside) the nucleus favours pinning,
because one saves pairing energy  placing the vortex inside the
nucleus.
Second, to build  a  $\nu=1$ vortex 
inside the nucleus
requires the formation of Cooper pairs out of single-particle levels of
opposite parity. 
This is strongly hindered by the spatial quantization
associated with finite size nuclear effects which leads
to a distribution of levels around the Fermi energy essentially displaying
either positive or negative parity (in other words, a very small $\nu=1
(\pi = (-1)^{\nu} = -1$) phase space \cite{footnote2}).
The consequences of spatial quantization on a $\nu=1$ vortex 
can be further clarified by comparing the corrispondent solution of
the De Gennes equations with that associated with a
$\nu=2$ vortex (see Figs. 3(c) and (d): note that within the 
present context the stability or less of this solution is immaterial). 
It is seen  that the pairing gap in the
case $\nu=2$ displays a 
quadratic, rather than a linear dependence on $\rho$, reaching 
the asymptotic value at a
larger distances from the $z-$axis than in the $\nu=1$ case. 
It can also be seen that the pairing gap in the case $\nu=2$ is 
modified very little by the presence of 
the nucleus.
This is because $\nu=2$ vortices thrieve on a subspace
of single-particle levels all displaying essentially the same parity.

\begin{figure}
\centering
\includegraphics{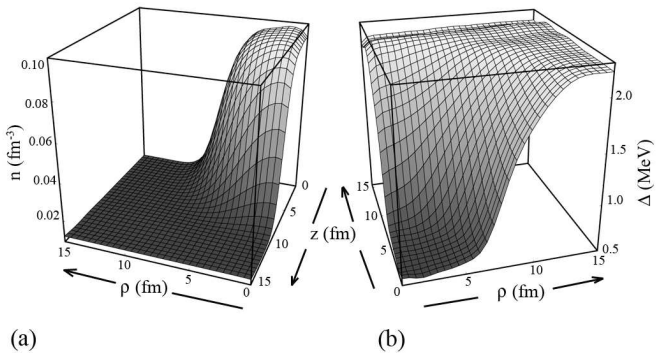}
\caption{
Density and pairing gap calculated for a nucleus immersed in the neutron sea, at
the Fermi energy $E_F = 5.8 $ MeV. They are spherically symmetric,
but are calculated in the cylindrical box described in the text (radius 30 fm
and  height 40 fm). Only the region $ 0 < z < 15 $ fm and $ 0 < \rho < 15 $ fm
is shown.}
\label{fig:figure1}
\end{figure}

\begin{figure}
\centering
\includegraphics{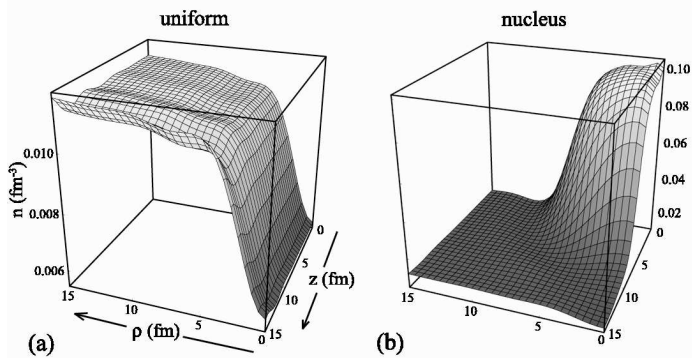}
\caption{
Density associated with a  $\nu=1$ vortex, calculated 
in the cylindrical box described in the text (radius $30 $ fm 
and height  45 fm), without and with a nucleus at the center of the box.
Only the region $ 0 < z < 15$  fm, $ 0 < \rho < $15 fm is shown.
(a) Density associated with a $\nu=1$ vortex in the cell without the nucleus.
(b) Density associated with a $\nu=1$ vortex in the presence of the nucleus.
}
\label{fig:figure2}
\end{figure}

\begin{figure}
\centering
\includegraphics{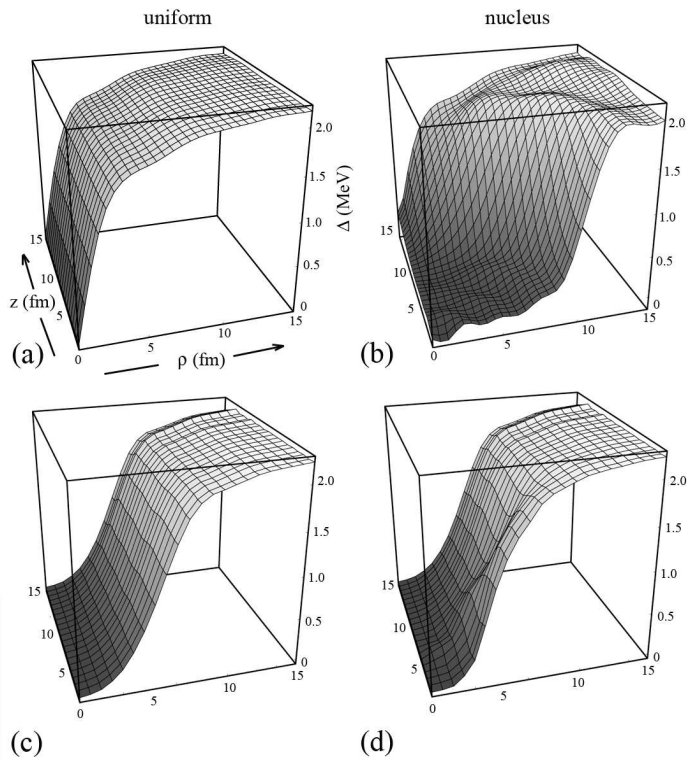}
\caption{
Pairing gap associated with  a $\nu=1$ and a $\nu=2$ vortex, calculated 
in the cylindrical box described in the text (radius $30 $ fm 
and height  40 fm), without and with a nucleus at the center of the box.
Only the region $ 0 < z <15$  fm, $ 0 < \rho < $15 fm is shown.
(a) Gap associated with a $\nu=1$ vortex in the cell without the nucleus.
(b) Gap associated with a $\nu=1$ vortex in the presence of the nucleus.
(c) Gap associated with a $\nu=2$ vortex in the cell without the nucleus.
(d) Gap associated with a $\nu=2$ vortex in the presence the nucleus.}
\label{fig:figure3}
\end{figure}

We shall now consider the energy associated with the various situations
shown above, for the case of  a $\nu=1$ vortex, and calculate the correspondent 
pinning energy. In our calculation,
the energy $E_{tot}$ of a given configuration  receives 
contributions from  three sources: 
the kinetic energy, the mean field potential (HF) energy, and the pairing energy,
so that 
$E_{tot}= E_{kin} + E_{pot} + E_{pair} $.
These contributions are displayed in Table 1
for the uniform case (no vortex, no nucleus in the cell: we call the value of 
the total
energy $E_U$),
for an interstitial vortex (no nucleus in the cell, $E_{I}$),
for an isolated nucleus (no vortex in the cell, $E_N$), and finally for a
vortex pinned on the nucleus ($E_{P}$).   
For the first two cases, our calculation is essentially 
the same as the one performed in ref. \cite{Bulgac}, although using a different 
pairing  interaction. 
The energy cost to create an interstitial vortex or a pinned vortex  are 
given by $E_{IU} = E_{I} -E_U$ and by $E_{PN} = E_{P} - E_N$, 
respectively \cite{footnote3}. 
 In order to be meaningful, the cost to create a vortex should refer 
to two systems  with the same number of neutrons moving in the same box.
The corresponding solution of the
self-consistent equations (1-4) leads to  a slight change $\delta
E_F$ of the
Fermi energy in going from a system without a vortex to a 
system with a vortex.
For an infinitely large box, the change tends to zero.
In the present calculation the change  was $\delta
E_F \approx  $ 0.05 MeV.

\begin{table}
\begin{tabular}{c|c|c|c|c|c|c} 
\hline
& & $E_{kin}$ & $E_{pot}$ & $  E_{pair}$ & $ E_{tot} $ \\ 
\hline
&Uniform & 6841.9 & -1735.3& -1322.1 & 3784.5 & \\
\hline
&Vortex, int.& 6776.0& -1737.5 & -1203.9& 3834.6&\\
\hline
&Nucleus & 9971.9 & -5784.0 & -1274.5 & 2913.4&\\
\hline
& Vortex, pinn. & 9893.4 & -5806.1 & -1120.5 & 2966.8&\\
\hline
\end{tabular}
\caption{
The total energy $E_{tot}$, subdivided into the three 
contributions arising from  the kinetic energy
$E_{kin}$, the potential energy $E_{pot}$ 
and the pairing energy $E_{pair}$, is shown  
for each of the four configurations discussed in the text
(Wigner cell without the nucleus, cell without the nucleus and 
with an interstitial vortex,  
cell with the nucleus, cell with the nucleus and a pinned  vortex).
 }
\end{table}

We  have evaluated the error associated with 
the finite mesh size used in integrating these 
equations ($\Delta \rho = \Delta z = 0.25$ fm).
In particular we have found that the value of the pinning energy, 
although  obtained from the subtraction of large numbers, is remarkably stable
with respect to changes in the box size. In fact we estimate that the
absolute value of the error associated with this quantity 
is less than 2 MeV.

The main contribution to $E_{ IU } = E_I - E_U$ 
as reported in Table 1  originates from 
the pairing  energy ($\Delta E_{pair}$= 118.2 MeV), 
and  is associated  with the 
decrease  of the pairing field in the region close to  the vortex axis 
(see Fig. 3(a)). At the same time, there is a reduction of kinetic energy
due to the reduced population of levels above the Fermi 
energy. On the other hand, one also expects a 
positive contribution to the kinetic energy in the presence of a vortex, 
associated  with the irrotational flow around the axis. In the present case 
the balance turns
out to be negative ($ \Delta E_{ kin } = -65.9$ MeV). 
The third contribution to $E_{IU}$ arises from a modification in the (HF)
mean field. 
This energy is related to the redistribution of 
the particles  around the vortex, which, in the case under discussion, 
is small 
($\Delta E_{pot} $= -2.2 MeV). 
Adding the three contributions leads to
$
E_{IU} =    \Delta E_{pair} +\Delta E_{kin}  + \Delta E_{pot} = 
118.2 - 65.9 -2.2 $ MeV = 50.1 MeV.
Analogous considerations can be made to calculate the energy cost $E_{PN}$ 
to create a vortex pinned on the nucleus. 
We have in this case 
$
E_{PN} =    \Delta E_{pair} +\Delta E_{kin}  + \Delta E_{mean} =
$ 
154.0 - 78.5 -22.1 MeV = 53.4 MeV. 
It is remarkable that the cost in pairing energy for a vortex  pinned on a nucleus 
(154.0
MeV) is larger  than the cost involved in the creation of  a vortex in uniform matter 
(118.2 MeV). 
One might have thought that because the vortex tends to lower
the pairing gap it would cost less pairing energy to create it 
in the  
region occupied by the nucleus, where the pairing gap  is lower than in uniform matter
(that is, far away from the nucleus, see Fig. 1(b)).  
However, 
because the vortex tends to avoid the nucleus, 
its presence affects the neutron pairing gap essentially only at  the nuclear 
surface. For example, for $z \approx 0, \rho
\approx $ 7 fm in the uniform case, the pairing gap changes from 
$\Delta_{unif}$ 
(2.2 MeV, $\nu =0$ situation )
to a value of 1.7 MeV ( $\nu=1$ vortex, cf. Fig. 3(a)).
In   the pinned case the pairing gap
changes from 1.6 MeV ($\nu =0$ situation) to about 0 MeV ($\nu=1$ vortex).  
In other words,
the main effect of a  vortex in the pinned situation is to
zeroth  the pairing gap at the nuclear surface,
an effect involving an important amount of pairing energy.
The pinning energy is defined as the difference between the energy cost 
to create a pinned vortex,
and to create a vortex in uniform matter \cite{epstein}. 
For the Fermi energy under consideration ($E_F$ = 5.8 MeV) it amounts to
$E_{PN} - E_{IU} $ = 3.3 MeV [13].

One can conclude that quantal finite size effects, in particular the spatial
quantization leading to shell structure, have important effects on the
vortex-nucleus interplay.


\begin{thebibliography}{9}
\bibitem{footnote2}  It could be argued that spin-orbit interaction, 
not considered in the present calculation, will shift the relative energies 
of levels of different parity 
around the Fermi energy, eventually reducing this effect.
However, the S=0 pairing field connects 
single-particle levels with antiparallel
spin projections, which are both either  shifted 
down by the spin-orbit interaction (for $l(1)>0 , s_z(1)>0; -(l-\nu)(2)<0, s_z(2)<0$)
or up (for $l(1)>0 , s_z(1)<0; -(l-\nu)(2)<0, s_z(2)>0$). 
One thus does not expect  a significant change of the phase space 
density relevant for vortex formation.
\bibitem{footnote3}To be noted that the energy cost to create a vortex  depends linearly on the height of 
the cylindrical box used for the calculation, and shows a logarithmic 
divergence with respect to the radius of the box. 
We have verified that our results obey this geometrical scaling.
Such dependences, however, disappear when we consider the  pinning energy,
which is independent of the box size for sufficiently large boxes.
\end{thebibliography}
\end{document}